\documentclass[preprint,12pt]{elsarticle}
\usepackage{setspace}
\usepackage[]{natbib}
\usepackage{graphicx,amssymb,amsmath,times}
\usepackage{subfigure}
\usepackage{lscape}
\journal{New Astronomy}
\begin{document}
\begin{frontmatter}
\title{Extremely High energy peaked BL Lac nature of the TeV blazar Mrk 501}
\author[label1,label2]{K K Singh\corref{corr}}
\ead{kksastro@barc.gov.in}
	\author[label1]{P J Meintjes} \author[label1]{F A Ramamonjisoa} \author[label2]{A Tolamatti} 
\address[label1]{Physics Department, University of the Free State, Bloemfontein- 9300, South Africa}
\address[label2]{Astrophysical Sciences Division, Bhabha Atomic Research Center, Mumbai- 400 085, India}
\begin{abstract}
Extremely High energy peaked BL Lac (EHBL) objects are a special class of blazars with peculiar observational properties 
at X-ray and $\gamma$--ray energies. The observations of these sources indicate hard X-ray and $\gamma$--ray spectra and 
absence of rapid flux variations in the multi-wavelength light curves. These observational features challenge the leptonic 
models for blazars due to unusually hard particle spectrum in the emission region of the blazar jet and provide a strong 
motivation for exploring  alternative scenarios to interpret the broad-band emission from blazars. At present, only few 
TeV blazars have been observed as EHBL objects in the extragalactic Universe. Due to their hard $\gamma$--ray spectra and 
long term variability, the observations of EHBL type of blazars at different redshifts help in probing the cosmic magnetic field and 
extragalactic background light in the Universe. Such objects also provide astrophysical sites to explore the particle acceleration 
mechanisms like magnetic reconnection and second order Fermi acceleration. Therefore, it has become important to identify 
more objects as EHBL using the observations available in the literature. Recent studies on the blazar Mrk 501 indicate 
that this source may exhibit an EHBL behaviour. In this paper, we use long term observations of Mrk 501 to explore its nature. 
Two sets of data, related to low and high/flaring activity states of Mrk 501, have been presented and compared with the observed 
features of a few well known EHBL type of blazars. We find that the spectral features of the blazar Mrk 501 indicate an EHBL nature 
of the source. Whereas, the temporal characteristics with fast variability during the high activity state of the source in X-ray 
and $\gamma$--ray energy bands are not compatible with the behaviour of EHBL type of blazars. However, Mrk 501 can be considered as 
an EHBL candidate in its low emission state. We also discuss the implications of identifying more EHBL objects using present 
and future ground-based $\gamma$-ray observatories.
\end{abstract}
\end{frontmatter}

\section{Introduction}  

Blazars are recognized as a special class of radio-loud active galactic nuclei having jets oriented along the line 
of sight of the observer at the Earth. The blazar jets are characterized as the relativistic  plasma outflows ejected 
from the supermassive black hole at the center of the host galaxies with an elliptical morphology \cite{Urry1995,Padovani2017}. 
The origin and launching of the relativistic jets have not been fully understood, however the Blandford \& Znajek process 
suggests that the jet is powered by the rotational energy of the black hole at the center \cite{Blandford1977}. 
Results from the multi-wavelength observations using various space and ground-based telescopes over the last three decades 
indicate that blazars are the dominant extragalactic sources of electromagnetic radiation from microwave/radio to 
very high energy (VHE; E $>$ 50 GeV) $\gamma$--rays. These objects are observed to exhibit high luminosity 
(up to 10$^{48}$-10$^{49}$~erg~s$^{-1}$) in the $\gamma$--ray energy band. The observed multi-wavelength emission from blazars 
is characterized as a non-thermal continuum, strongly polarized and highly variable at timescales from minutes to 
years \cite{Aharonian2007,Ackermann2016,Singh2018}. The dominant non-thermal emission from blazars is observed to originate 
from the relativistic jets. The small viewing angle of the jet makes blazar emission strongly beamed and Doppler boosted 
which dominates the observed broad-band spectral energy distribution (SED) of these sources. In the simplest scenario, the emission 
region is assumed to be relativistically moving along the jet axis with a bulk Lorentz factor same as the Doppler factor ($\delta$) 
in the small viewing angle ($\le 10^{\circ}$) approximation. Therefore, the observed bolometric luminosity of blazars is Doppler 
boosted by a factor of $\delta^4$. The observed frequency of photons is amplified by $\delta$ while the measured timescales are 
reduced by $1/\delta$. These simple arguments are supported by the measurements of extremely high bolometric luminosities 
varying at very short timescales. 
\par
The SED of blazars is characterized by a two-hump structure peaking at low and high energies. The origin of the first hump peaking at 
lower energies ranging from IR/optical to X-rays is attributed to the synchrotron radiation from the relativistic electrons in the 
jet magnetic field. The observations of high degree of polarization at radio and optical wavebands from blazars also support the 
hypothesis that synchrotron is the dominant process at low energies. Additionally, this also supports the idea of highly 
ordered magnetic field in the jet. The physical process involved in the second hump peaking at energies from high energy 
(HE; E $>$ 100 MeV) to VHE $\gamma$--rays is not completely known and has become a subject of intense research in the field of 
blazar astrophysics today. However, two alternative scenarios namely \emph{Leptonic} and \emph{Hadronic} are considered for 
modelling the $\gamma$--ray emission from blazars. In the Leptonic scenario, the emission of $\gamma$--ray photons 
is attributed to the inverse Compton (IC) scattering of the low energy target photons by the relativistic leptons 
(electrons and positrons)  in the blazar jet. Two different physical processes namely synchrotron self Compton (SSC) and external 
Compton (EC) are used to explain the $\gamma$--ray emission from blazars within the Leptonic scenario depending on the dominance 
of target photon field. If the low energy synchrotron photons scatter mainly off the relativistic electrons 
(which produce the synchrotron radiation) and gain energy to become $\gamma$--ray photons, the model is referred to 
as SSC \cite{Maraschi1992,Bottcher2007,Abdo2010}. In the case of an EC model, the target photon fields originating from regions 
(accretion disk, broad-line region, dusty torus, illuminated molecular clouds, and cosmic microwave background radiation) external to the 
jet dominate over the synchrotron photons for IC scattering \cite{Dermer1992,Sikora1994,Agudo2011}. On the other hand, in the 
Hadronic scenario, the $\gamma$--ray emission is explained by considering the presence of relativistic protons in the blazar jet. 
The processes like proton synchrotron and photo-hadronic interactions followed by subsequent pion decay and secondary particles 
cascade are invoked to describe the observed $\gamma$--ray emission from the blazars \cite{Mannheim1993,Aharonian2002,Bottcher2009}. 
Alternatively, the hybrid scenario involving lepto-hadronic processes for $\gamma$--ray emission has also been proposed in the literature 
\cite{Bottcher2013,Cerruti2015,Petropoulou2015}. In the hybrid scenario, the hadronic processes associated with relativistic protons 
are assumed to significantly contribute to the emission of observed $\gamma$--rays along with the IC process by the relativistic electrons 
in the blazar jet \cite{Yan2015}. The information related to the exact content of protons in the jet is very important for understanding the 
jet dynamics and launching from the center of the blazar host galaxy. The acceleration processes for electrons and protons up to relativistic 
energies in the blazar jet are not exactly known.   
\par
Single zone leptonic models with the given population of relativistic electrons and positrons in the steady state have been 
successfully used to explain the non-thermal simultaneous or quasi-simultaneous multi-wavelength emission from almost all the
observed blazars \cite{Bottcher2013,Celotti2008,Ghisellini2010}. However, the detection of a short term variability at minutes 
scale challenges the one zone leptonic scenario because it requires very high values of the bulk Lorentz factor for the 
emission region \cite{Begelman2008,Barkov2012,Aharonian2017}. Time dependent leptonic models with single emission zone 
are also being used to explain the variability of blazars in the literature \cite{Joshi2011,Mastichiadis2013,Singh2017}.
Several complex leptonic models with multi-zone emissions in the jet have been proposed  to understand the short 
timescale $\gamma$--ray variability observed from many blazars \citep{Georganopoulos2003,Tavecchio2008,Giannios2010}.    
The detection of statistically significant neutrino events during the high activity from the same source in $\gamma$--rays 
would provide unique tools to discriminate between the leptonic and hadronic components in the $\gamma$--ray emission 
from blazars. However, recent detection of TeV neutrino events with low statistical significance (less than 5$\sigma$) 
from the blazars TXS 0506+056 \citep{IceCube2018} poses a serious challenge to the dominance of hadronic processes in blazar 
jets for $\gamma$--ray emission and reinforces the lepto-hadronic scenario. The detection of a special class of blazars  
known as \emph{Extremely High energy peaked BL Lac} objects (EHBLs) having an exceptionally hard TeV spectrum with steady 
emission over a long timescale challenges the leptonic models for the blazar emission. These observations indicate that exploring 
the exact physical process for $\gamma$--ray emission from blazar jets is very exciting and an open problem in high energy astrophysics. 
\par  
In this paper, we use observed properties/parameters of a well-studied blazar Mrk 501 available from the  long term  
observations to explore its nature during different activity states of the source. In Section 2, we briefly describe different 
types of blazars based on their observational properties so far. Some of the important characteristics of EHBL blazars are discussed 
in Section 3. In Section 4, we use all the observations available on Mrk 501 in the literature to study the behaviour of X-ray and 
$\gamma$--ray emission from the source and discuss its possible association with the well known EHBLs observed to date. Finally we summarize 
the findings of our study in Section 5. The $\Lambda$CDM cosmological model with parameters $\Omega_m=0.27$, $\Omega_{\Lambda}=0.73$ and 
$H_0=70$~km~s$^{-1}$~Mpc$^{-1}$ has been adopted in the paper.

\section{Blazars Types}
Although blazars in general have similar emission properties, they are classified in different types based on their observed characteristics. 
The widely accepted different types for blazars are briefly described below: 

\subsection{Optical types}

On the basis of the observed characteristic optical spectral lines and their rest frame equivalent width (EW), blazars have been 
classified in two types: BL Lacertae (BL Lac) objects and Flat Spectrum Radio Quasars (FSRQs) \cite{Urry1995,Padovani2017,Stocke1991}. 
BL Lac objects exhibit no or weak spectral lines (EW $<$ 5~\AA) against the bright featureless optical continuum whereas 
FSRQs have prominent broad emission lines (EW $>$ 5~\AA). The observations of strong optical spectral lines from the FSRQs indicate the 
presence of broad-line emission regions illuminated by the bright accretion disk in this type of blazars. The accretion disk in FSRQs is
radiatively efficient with a near-Eddington accretion whereas BL Lacs have less radiatively efficient accretion disks with a sub-Eddington 
accretion. Due to the presence of a bright accretion disk, FSRQs are observed to be more luminous than BL Lac class and it provides intense 
external target photon fields for IC scattering that produces $\gamma$--ray photons in the jet emission region of the FSRQs in the Leptonic scenario. 
Also, EC models are observed to be more relevant for FSRQs than BL Lac type of blazars where SSC processes are dominant for $\gamma$--ray 
emission under the framework of the Leptonic scenario. This observational classification of blazars based on the optical emission is very simple 
and provides limited physical distinction for a large number of sources observed in different energy bands.

\subsection{SED types} 

In blazars, the non-thermal jet emission is found to be dominating at all energies from radio to $\gamma$--rays and the broad-band SED is very 
important to understand the emission process. The physical process for the origin of the low energy hump in the blazar SED is ascribed 
to the Doppler boosted synchrotron radiation produced by the relativistic leptons in the jet magnetic field. Therefore, the energy 
corresponding to the peak position of the low energy hump (E$_{syn}^p$) in the observed SED is used as an important parameter for the 
classification of blazars. From the observations, it is found that the rest frame peak energy of the low energy hump 
follows a distribution in the waveband from IR/optical to X-rays for BL Lac type of blazars. Therefore, on the basis of rest frame peak energy, 
BL Lac objects have been classified in three classes namely \cite{Padovani1995}: Low energy peaked BL Lac (LBL: E$_{syn}^p \le$ 0.5 eV), 
Intermediate energy peaked BL Lac (IBL: 0.5 eV $\le E_{syn}^p \le$ 5 eV) and High energy peaked BL Lac (HBL: E$_{syn}^p \ge$ 5 eV). The low energy 
hump in the SED of FSRQs is observed to peak at  E$_{syn}^p$ $\le$ 0.5 eV. Hence, FSRQs can be associated with the LBL class of objects under 
the classification scheme based on the blazar SED. An extension for the general classification of blazars on the basis of the position of the rest 
frame synchrotron peak frequency ($\nu_{syn}^p$) in the SED is also proposed by \cite{Abdo2010}. According to this classification, all blazars 
are classified as: Low Synchrotron Peaked (LSP: $\nu_{syn}^p \le 10^{14}$ Hz), Intermediate Synchrotron 
Peaked (ISP: 10$^{14} \le \nu_{syn}^p \le 10^{15}$ Hz) and High Synchrotron Peaked (HSP: $\nu_{syn}^p \ge 10^{15}$ Hz). Most of the blazars detected 
so far at TeV energies belong to the HBL or HSP class. A small fraction of HBLs is observed to exhibit a SED with the position of the synchrotron peak at 
hard X-ray energies (E$_{syn}^p \ge$ 5 keV) or higher (E$_{syn}^p \sim$ MeV-GeV). These type of blazars are referred to as EHBLs 
and constitute an interesting class of sources to be explored in the extragalactic Universe at X-ray and $\gamma$--ray 
energies \cite{Costamante2001}.  

\subsection{Blazar Sequence}

The observed values of the synchrotron peak frequency ($\nu_{syn}^p$) vary between $\sim$ 10$^{12}$ Hz to over 10$^{18}$ Hz for all type of 
blazars \cite{Giommi2012}. A strong anti-correlation obtained between the bolometric luminosity and the rest frame synchrotron peak frequency 
forms a sequence referred to as the \emph{Blazar Sequence} \cite{Ghisellini2017}. According to the observed \emph{Blazar Sequence}, FSRQs 
(having the lowest value of $\nu_{syn}^p$) have the highest luminosity whereas HBLs (having the highest value of $\nu_{syn}^p$) have the lowest 
luminosity. This indicates that FSRQs and BL Lacs are high and low-power blazars respectively. The Compton dominance (ratio of the peak luminosity 
in high energy component to the peak luminosity in synchrotron or low energy hump in the SED) for all types of blazars decreases with increasing 
synchrotron peak frequency. It means the Compton dominance is maximum for FSRQs and minimum for HBLs. Also, the high energy component of SED peaks 
at MeV-GeV energies for FSRQs and at TeV energies for HBLs. For EHBL type of blazars, the IC peak is observed at higher energies than the classical 
HBLs because of their harder $\gamma$--ray spectra. Therefore, it is very interesting to study the Compton dominance for EHBL type of blazars which 
will help to understand the strong radiative cooling of the relativistic leptons due to IC scattering in blazars.

\section{Extremely High energy peaked BL Lac (EHBL) blazars}

The existence of EHBL type of blazars with hard power law spectrum and E$_{syn}^p$ up to 100 keV was first predicted by 
\cite{Costamante2001} using X-ray observation of blazars by the \emph{BeppoSAX}. These sources are suspected to be hidden in 
the HBL class of blazars. Some of the key theoretical predictions about EHBL type of blazars are as follows:

\begin{itemize}

\item Possible candidates for invoking hadronic models with hard particle spectrum and strong magnetic field (more than 100 G) to explain 
      the observed high energy $\gamma$--ray emission and long timescale variability \citep{Cerruti2015}.

\item With the lowest intrinsic luminosity and highest synchrotron peak energy, these objects follow the blazar sequence as 
      FSRQ$\rightarrow$LBL$\rightarrow$IBL$\rightarrow$HBL$\rightarrow$EHBL (in the order of decreasing luminosity and increasing synchrotron 
      peak energy).

\item The peak energy in the synchrotron hump indicates the maximum energy of the particles in the emission region. This indicates that EHBL 
      jets are the extreme and efficient astrophysical accelerators in the Universe. However, the very hard spectrum of the injected population 
      of relativistic particles with index less than 1.5 challenges the simple acceleration models like diffusive shock acceleration in the 
      blazar jets.

\end{itemize}

Since their prediction, the study of EHBL type of blazars has become an intense area of research in multi-wavelength astrophysics. 
Some of the observable properties of EHBLs are summarized below:

\begin{itemize}

\item Difficult to be identified by radio observations because of weak or faint radio emission and are mostly observed at X-ray and 
      TeV $\gamma$--ray energies unlike any other class of blazars. 
		
\item Low and high energy peaks in the broad-band SED are shifted to hard X-ray and VHE $\gamma$--rays.
			
\item IR and optical emissions are dominated by the thermal emission from the host galaxy and non-thermal jet emission dominates at 
      UV energies. 			

\item Exhibit exceptionally hard intrinsic X-ray and TeV $\gamma$--ray spectra (spectral index $\leq$ 2) and lack variability at short 
      timescales (variability timescale of several days is observed) despite the predictions from leptonic models. 
			
\item Pose a strong challenge for the standard one zone leptonic SSC models for high energy emission because they require large values of 
      the Doppler factor ($\delta >$ 50) and very high values of the minimum Lorentz factor ($\gamma_{min} >$ 150) for electrons.

\end{itemize}  

From the above discussions it is obvious that the observed emission properties of EHBLs are very different from the classical BL Lac type of 
blazars and HBLs in particular. Therefore, it will not be appropriate to use the source parameters derived from the study of BL Lac objects as 
a probe for EHBLs. However, these objects with very peculiar observational properties constitute an interesting class of blazars for exploring 
the broad-band emission processes. A very small population of EHBLs detected so far gives a strong motivation for the identification of new sources 
belonging to this subclass of blazars. A brief review of the observational properties of the VHE $\gamma$--ray emission from the prominent EHBL 
type of blazars detected to date is given in the Appendix. 

\section{EHBL nature of Mrk 501}
Mrk 501 was first presented as a spheroidal galaxy with faint corona and star-like nucleus in the 5th list of galaxies with UV continuum in 1972 
from the sky survey during 1969--70 at Byurakan Observatory \cite{Markaryan1972}. In 1975, this object was identified as an elliptical galaxy 
B2 1652+39 (Mrk 501) at redshift $z$ = 0.0337 from the spectrographic, photometric, polarimetric and radio observations \cite{Ulrich1975}. 
The dominant contribution of non-thermal emission to the luminosity of the nuclear region was also detected from B2 1652+39 with optical 
properties similar to the elliptical galaxies and BL Lac objects. In 1981, this source was reported in a catalog of radio sources with 
flux density above 1 Jy at 5 GHz \cite{Kuehr1981}. The redshift of Mrk 501 was again measured to be $z$ = 0.034 from the stellar 
absorption/emission lines of the host galaxy using direct imaging and spectroscopic observations of selected BL Lac objects from the 
1 Jy radio catalog performed during 1985--1990 \cite{Stickel1993}. 
The first VHE $\gamma$--ray emission from this source above 300 GeV was discovered by the Whipple telescope in 1995 \cite{Quinn1996} 
soon after the discovery of TeV $\gamma$--ray photons from the first extragalactic source Mrk 421 in 1991 \cite{Punch1992}. Since the 
discovery of TeV $\gamma$--ray emission from Mrk 501, several intensive broad-band studies have been performed on this blazar, but 
its exact nature has not yet been completely understood.
\par
In the following, we use  all the observational results on the blazar Mrk 501 which are available in the literature today and compare 
them with the properties of well known EHBLs (Appendix) to characterize its nature. Among the various characteristics of the EHBL type 
of blazars as known so far, an object can be characterized as an EHBL if it exhibits important observational features like: 
(i) hard intrinsic VHE spectrum with a spectral index $\Gamma_{int} \leq$ 2 and peak energy in the second hump of the SED above 1 TeV, 
(ii) hard \emph{Fermi}-LAT spectrum with a photon spectral index $\Gamma_{LAT} \leq$ 2 in the MeV-GeV energy band, (iii) hard X-ray spectrum 
with a photon spectral index $\Gamma_X \leq$ 2 and synchrotron peak energy above E$_{syn}^p \sim$ 10 keV, and (iv) no short term variability 
in the TeV $\gamma$--ray emission.

\subsection{Intrinsic VHE or TeV $\gamma$--ray spectrum}
After its identification as a TeV $\gamma$--ray source in 1996, Mrk 501 has become a prominent target for VHE observations by ground-based 
$\gamma$--ray telescopes until today. Assuming that the physical process for TeV $\gamma$--ray production is similar for a given class 
of blazars like EHBLs, the differences in the observed TeV spectra of sources at different redshifts can be attributed to the partial 
absorption of VHE photons due to the interaction with low energy extragalactic background light (EBL) photons. EBL represents the repository 
of the photons in the wavelength range from far-IR to optical/UV produced by the stars and galaxies throughout the history of the Universe. 
The VHE $\gamma$--ray photons interact with low energy EBL photons via photon-photon pair production while propagating from source to the 
observer and are absorbed in the intergalactic space. This absorption leads to the energy and redshift dependent attenuation of the 
VHE $\gamma$--ray photons emitted from the source  which is characterized by a physical quantity called \emph{optical depth} ($\tau$). 
The detailed procedure for the estimation of $\tau$ as a function of the source redshift ($z$) and energy of the VHE photons ($E$) 
for different models of EBL photon density is discussed in \cite{Singh2014}. The observed ($F_{obs}$) and intrinsic ($F_{int}$) VHE 
spectra of a source at cosmological distance are related as
\begin{equation}
			F_{obs}=F_{int}~~e^{-\tau(E,z)}
\end{equation}
where e$^{-\tau}$ is referred to as \emph{EBL attenuation factor}. This energy dependent attenuation makes the observed VHE $\gamma$--ray spectrum 
softer than the emitted intrinsic spectrum of the source. If $F_{obs}$ and $F_{int}$ are described by a power law with photon spectral indices 
$\Gamma_{obs}$ and $\Gamma_{int}$ respectively, then it is expected that $\Gamma_{int}~ < ~\Gamma_{obs}$ for a given source. If the TeV $\gamma$--rays 
are produced by the IC scattering in the Klein-Nishina (KN) regime, the intrinsic VHE spectrum is softer than the Thomson regime \cite{Sahayanathan2012}. 
However, the hard VHE spectra of the EHBL type of blazars contradict the leptonic origin and suggest the hadronic origin of the TeV photons from these 
sources. We have used the most recent and updated EBL model proposed by Franceschini \& Rodighiero \cite{Franceschini2017} to estimate the optical depth 
($\tau$) for $\gamma$--ray photons in the energy range 50 GeV--15 TeV emitted from the sources at different redshifts in the range $z$ = 0.034-0.230. 
The corresponding \emph{EBL attenuation factor} as a function of energy for the EHBLs described in the Appendix is shown in Figure \ref{fig:Fig1}. 
It is evident that the EBL absorption effect on the TeV $\gamma$--ray spectra of Mrk 501 is minimal as compared to the other EHBLs reported 
in Figure \ref{fig:Fig1}. Therefore, the observed TeV spectra of Mrk 501 can be approximated as the intrinsic emission spectra which reflect the best 
physical process for the production of TeV $\gamma$--ray photons at the source. 
\begin{figure}
\begin{center}
\includegraphics[scale=0.62,angle=-90]{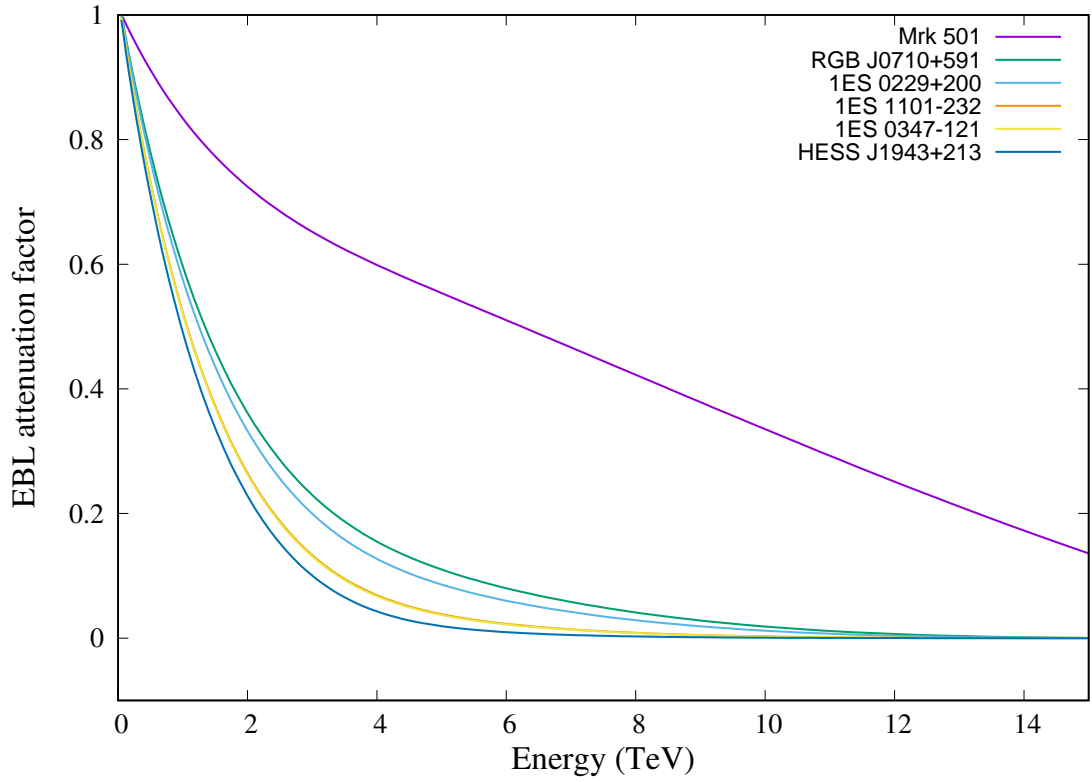}
	\caption{EBL attenuation factor (e$^{-\tau}$) as a function of TeV $\gamma$--ray photon energy for well known EHBLs (Appendix) at 
	        different redshifts ($z$ = 0.125-0.230) and for the blazar Mrk 501 at $z$ = 0.034. The EBL model described in \cite{Franceschini2017} 
		is used to estimate the optical depth ($\tau$) for a given energy of VHE $\gamma$--ray photon ($E$) emitted from a source at 
		redshift ($z$).}  
\label{fig:Fig1}
\end{center}
\end{figure}
\par
\begin{figure}
\begin{center}
\includegraphics[scale=0.62,angle=-90]{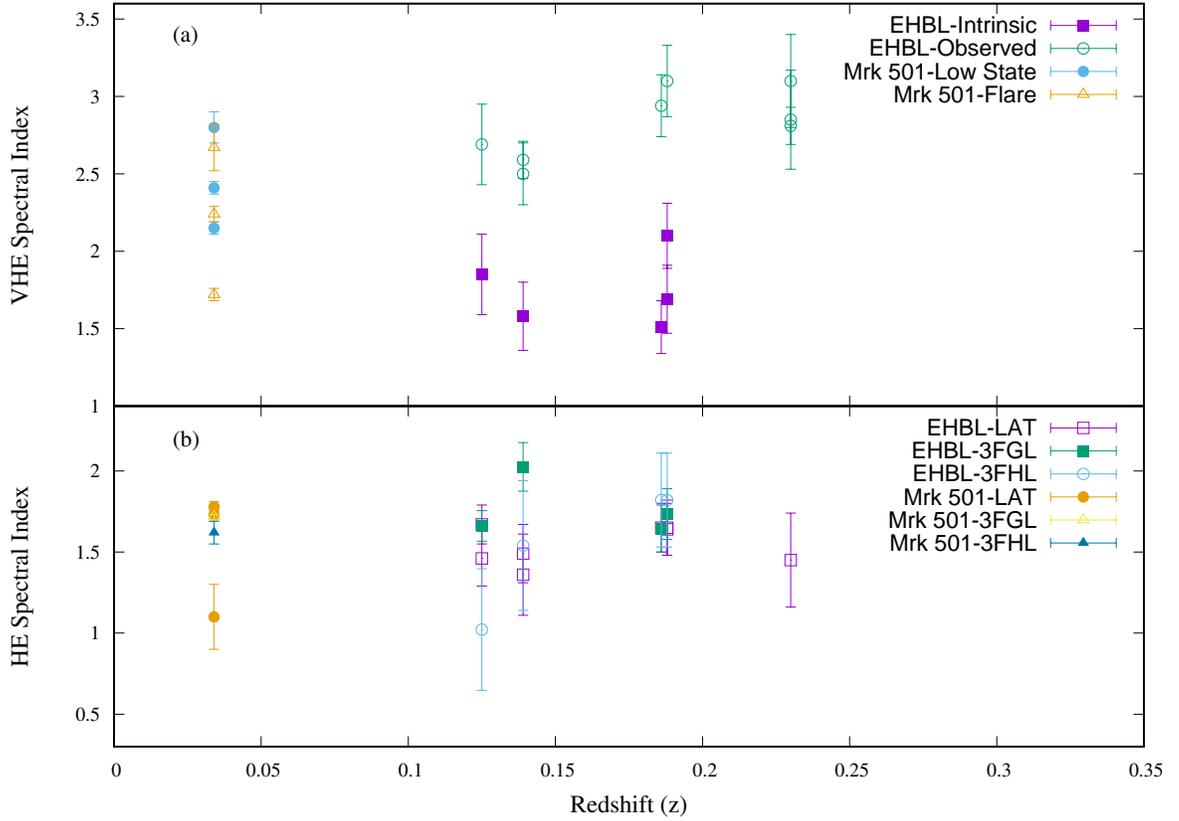}
	\caption{Comparison of power law spectral indices of Mrk 501 in VHE ($E >$ 50 GeV) and HE ($E >$ 100 MeV) $\gamma$--ray bands with the 
	well known EHBLs (Appendix) as observed so far. (a) VHE spectral indices are used from the literature based on observations using various 
	ground-based atmospheric Cherenkov telescopes like Whipple, HEGRA, MAGIC, VERITAS, H.E.S.S. and TACTIC reported in the TeV catalog. 
	(b) HE spectral indices have been obtained from the \emph{Fermi}-LAT observations. For Mrk 501, only maximum, minimum and average values 
	of the spectral indices are shown.}  
\label{fig:Fig2}
\end{center}
\end{figure}

The distribution of the power law spectral indices ($\Gamma$) in VHE $\gamma$--ray band for Mrk 501 and the well known EHBLs is shown in 
Figure \ref{fig:Fig2}(a). All the spectral indices values have been obtained from the literature reported in the online 
TeV catalog\footnote{http://tevcat.uchicago.edu}. For Mrk 501, we have used the spectral index measurements available in the literature 
since the discovery of the VHE $\gamma$--ray emission from this source in 1996 until today. The VHE observations of Mrk 501 are divided into 
two categories namely low and high/flaring states. In the low state, the VHE $\gamma$--ray emission from this source is found to be 
described by an average power law spectral index of 2.41$\pm$0.05 whereas in the flaring state the average spectral index is 2.24$\pm$0.06. 
This indicates that the spectral behaviour of VHE $\gamma$--ray emission from Mrk 501 does not change significantly during the flaring 
states of the source. However, an \emph{harder-when-brighter} behaviour, an observational characteristics of TeV blazars, has been observed 
for Mrk 501 at several occasions during its flaring activity in VHE regime \cite{Djannati1999,Albert2007}.
In Figure \ref{fig:Fig2}(a), the intrinsic spectral indices of the well known EHBLs (Appendix) differ significantly from their observed values 
indicating the effect of EBL absorption due to relatively large redshift of the sources in the range $z$ = 0.125-0.230. However, the values of VHE 
spectral indices for Mrk 501 are found to be consistent with the intrinsic spectral indices for EHBLs (EHBL-Intrinsic) within the error bars. 
This implies that the observed TeV $\gamma$--ray emission from Mrk 501 exhibits similar spectral characteristics as the EHBLs within the 
statistical uncertainty. 
\subsection{HE or \emph{Fermi}-LAT spectrum}   
The Large Area Telescope on board the \emph{Fermi Gamma-ray Space Telescope} satellite (\emph{Fermi}-LAT) provides HE $\gamma$--ray observations 
of the blazars in the energy range 100 MeV to more than 500 GeV \cite{Atwood2009}. The distribution of HE spectral indices derived from the 
\emph{Fermi}-LAT observations of Mrk 501 and other EHBLs (Appendix) is shown in Figure \ref{fig:Fig2}(b). For Mrk 501, HE $\gamma$--ray emission is 
observed to follow a power law distribution with an average photon spectral index of 1.73$\pm$0.03 in the energy range 0.1-300 GeV. This is 
consistent with the LAT spectral index for a sample of EHBLs (EHBL-LAT) as shown in Figure \ref{fig:Fig2}(b) and reported in \cite{Costamante2018}.
We have also used the results from the long term \emph{Fermi}-LAT observations on Mrk 501 and EHBLs reported in the last released LAT catalogs 
namely the Third \emph{Fermi}-LAT catalog (3FGL) and the Third catalog of hard \emph{Fermi}-LAT sources (3FHL) in Figure \ref{fig:Fig2}(b). 
3FGL catalog is based on the data in the energy range 0.1--300 GeV from the first four years of \emph{Fermi}-LAT observations \cite{Acero2015} 
and 3FHL includes the first seven years of data in the 10 GeV--2 TeV energy range \cite{Ajello2017}.  The values of LAT photon spectral index 
1.716$\pm$0.016 and 1.61$\pm$0.07 for Mrk 501 reported in the 3FGL and 3FHL catalogs respectively are consistent with the corresponding values 
for EHBLs within statistical uncertainties. The effect of EBL absorption on the \emph{Fermi}-LAT spectrum of the sources measured in the energy 
range 0.1-500 GeV is negligible for redshift $z \leq$ 0.5. Therefore, the LAT spectral index can be approximated as the intrinsic HE $\gamma$--ray 
index of the sources with the redshift $z \leq$ 0.5. In the present study, the values of LAT spectral index or HE $\gamma$--ray index 
are observed to be less than 2 for Mrk 501 and other well known EHBLs (Appendix). This implies that the IC or $\gamma$--ray peak will be 
shifted at TeV energies in the broad-band SED of these sources in log($E$) versus log($E^2$d$N$/d$E$) plane. Therefore, the HE spectral 
characteristics of Mrk 501 from the \emph{Fermi}-LAT observations support the EHBL nature of the source.
\begin{figure}
\begin{center}
\includegraphics[scale=0.62,angle=-90]{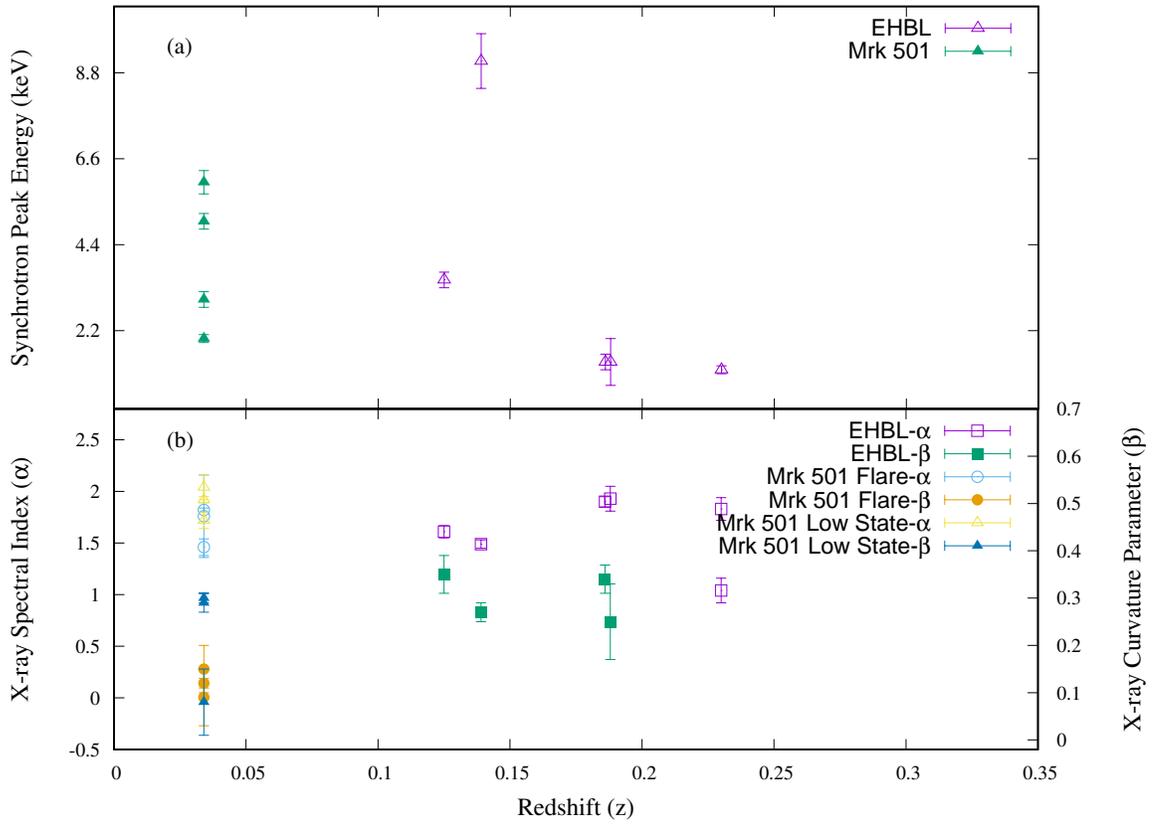}
	\caption{Comparison of X-ray emission parameters observed from Mrk 501 with the EHBLs (Appendix) as observed so far. 
	 (a) Synchrotron peak energy obtained from various X-ray observations performed in the energy range 0.3-79 keV 
	 (b) Spectral parameters obtained from the X-ray observations. Only maximum, minimum and average values of the spectral parameters 
	 are shown for Mrk 501.}  
\label{fig:Fig3}
\end{center}
\end{figure}

\subsection{X-ray emission characteristics}
The origin of X-ray emission from blazars has been properly understood and is ascribed to the synchrotron radiation emitted from the relativistic 
population of leptons (electrons and positrons) in the jet magnetic field. Therefore, the observed characteristics of the X-ray emission of blazars 
are widely used to characterize different types of blazars. We have used some of the important features like synchrotron peak energy and 
X-ray spectral index derived from the simultaneous observations with \emph{NuSTAR} and \emph{Neil Gehrels Swift Observatory} of Mrk 501 
and other EHBLs (Appendix) in the present study. The distribution of these characteristic parameters is depicted in 
Figure \ref{fig:Fig3}(a-b). The synchrotron 
peak energy (E$_{syn}^p$) shown in Figure \ref{fig:Fig3}(a) for different EHBLs has been obtained from the literature \cite{Costamante2018}. 
Detailed broad-band studies on Mrk 501 suggest a synchrotron peak energy between 2-5 keV in low or flaring state of the source 
\cite{Abdo2011,Furniss2015}. However, during the correlated X-ray and TeV flaring activity of Mrk 501 in 1997, the synchrotron peak energy 
was observed to be $\sim$ 100 keV with a shift of two orders of magnitude with respect to the low activity state \cite{Pian1998}. The average 
synchrotron emission from Mrk 501 is observed to peak in the energy range 2--100 keV \cite{Nieppola2006}. The position of the synchrotron peak in 
the broad-band SED of Mrk 501 is shifted to higher energies during the flaring state than that for the low emission state of the source. 
The X-ray emission from the sample of EHBLs (Appendix) is found to be described by a log-parabolic model with a photon spectral index ($\alpha$) 
and curvature parameter ($\beta$) in the energy range 0.3-79 keV \cite{Costamante2018,Wang2018}. The distribution of the spectral parameters 
$\alpha$ and $\beta$ for all the EHBLs along with Mrk 501 is shown in Figure \ref{fig:Fig3}(b). It is obvious from the Figure \ref{fig:Fig3}(b) 
that the X-ray spectral parameters of Mrk 501 are consistent with the corresponding parameters for EHBLs. However, during one of its flaring events, 
the X-ray emission from Mrk 501 was described by a power law with the hardest spectral index of 1.2$\pm$0.1 \cite{Pian1998}. A significant variation in 
the X-ray spectrum has been observed during the flaring activity of Mrk 501 with an \emph{harder-when-brighter} trend with respect to previous studies 
of the source \cite{Djannati1999,Abdo2011,Kapanadze2017}. Thus, the X-ray emissions of Mrk 501 show almost similar features like EHBLs during the 
flaring state of the source.

\begin{figure}
\begin{center}
\includegraphics[scale=0.62,angle=-90]{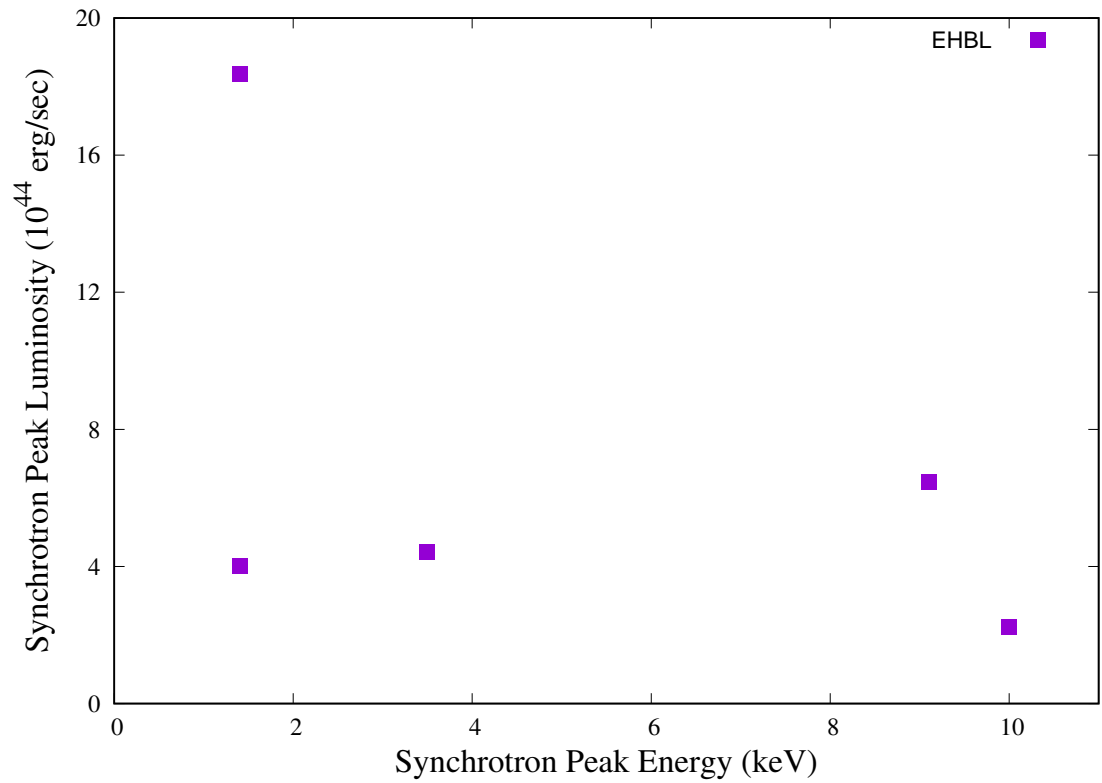}
\caption{Synchrotron peak luminosity ($L_{syn}^p$) as a function of synchrotron peak energy for EHBLs discussed in the Appendix.}  
\label{fig:Fig4}
\end{center}
\end{figure}

\subsection{Synchrotron peak luminosity}
From the prediction of the blazar sequence, EHBLs are expected to be least luminous. We have estimated the synchrotron peak luminosity using the 
results from the X-ray observations reported in \cite{Costamante2018} for the sample of EHBLs discussed in the Appendix. The synchrotron peak 
luminosity ($L_{syn}^p$) has been calculated using the formula:
\begin{equation}
	L_{syn}^p=\frac{F_p~\times~4\pi d_L^2}{(1+z)^{2-\alpha}}
\end{equation}
where $F_p$ is the synchrotron peak energy flux, $d_L$ is the luminosity distance and $\alpha$ is the observed photon spectral index. 
The distribution of the synchrotron peak luminosity as a function of the synchrotron peak energy for the EHBLs is presented in Figure 
\ref{fig:Fig4}. The synchrotron peak luminosity for the known EHBLs is obtained between 10$^{44}$-10$^{45}$~erg~sec$^{-1}$ which 
is consistent with the values estimated for blazars in the range 10$^{43}$-10$^{47}$~erg~sec$^{-1}$ \cite{Xiong2015,Ding2017}. 
The values of the synchrotron peak luminosity ($\sim 10^{45}$~erg~sec$^{-1}$) for Mrk 501 reported in the literature are in agreement with 
the values estimated for the EHBLs in the present study \cite{Abdo2011,Kapanadze2017}. During the flaring state of Mrk 501, the synchrotron peak 
position is shifted towards higher energy with a decrease in the synchrotron peak luminosity \cite{Pian1998}. We observe that Mrk 501 follows 
the blazar sequence like other EHBL type of blazars. However, more X-ray observations of the EHBLs in wide energy range are required to test 
the blazar sequence hypothesis with high confidence. 

\subsection{Variability}
Mrk 501 is one of the blazars known for its strong variability in all energy bands. The X-ray and VHE $\gamma$--ray emissions from Mrk 501 are observed 
to be highly variable during the flaring activity. In 2005, VHE $\gamma$--ray flux from this source  was observed to vary on a minute scale during 
a flare \cite{Albert2007}. The high activity state of Mrk 501 in X-ray and TeV $\gamma$--ray bands observed in June 2014 was characterized by a 
minimum variability timescale of a few minutes \cite{Chakraborty2015,Singh2019}. The variability characteristics from Mrk 501 contradict the 
hypothesis for temporal behaviour of the well known EHBLs (Appendix) where no short term variability is present. Therefore, the temporal characteristics 
of Mrk 501 observed in the X-ray and $\gamma$--ray energy bands are not compatible with the nature of EHBLs as observed so far. Therefore, it is very 
important to explore the temporal behaviour of more objects which are probable candidates for EHBLs using the long and short term observations available 
with the X-ray and $\gamma$--ray telescopes as the long term variability is expected to be one of the characteristics of such type of blazars.

\section{Summary}
In this work, we have explored the nature of the well studied blazar Mrk 501 which is frequently observed in high activity states at X-ray 
and $\gamma$--ray energies. We have used the X-ray and $\gamma$--ray observational results on Mrk 501 available in the literature to probe 
its nature as an EHBL. We find that the spectral characteristics of Mrk 501 measured in X-ray and $\gamma$--ray energy bands are broadly 
consistent with the behaviour of the well known EHBLs as reported in the literature so far. However, the temporal features with high variability at 
X-ray and VHE $\gamma$--ray energies during the flaring state of Mrk 501 are not in agreement with the nature of EHBLs where no variability or 
long term variability is expected to be present. Long term observations in the low activity state of Mrk 501 suggest that the source is a strong 
EHBL candidate in its low emission state. Recently, Ahnen et al. (2018) have concluded that the EHBL nature of Mrk 501 is a dynamic characteristic 
of the source which changes over time \cite{Ahnen2018}. In this study, the authors have used an extensive multi-wavelength observation on Mrk 501 
during March-July 2012 to characterize the broad-band emission from the source. It has also been suggested by the authors that the classification 
of blazars as EHBLs on the basis of observations has some diversity. Therefore, a comprehensive multi-wavelength study of the large sample of blazars 
in the redshift range 0.01$\leq z \leq$ 1 is very important to fully explore the nature and population of the EHBL type of blazars in the 
extragalactic Universe.
\par
The EHBL type of blazars is very important for probing the density of EBL photons because of the increased photon 
 statistics at the highest energy end of the spectrum due to hard VHE intrinsic spectral index. The TeV observations 
 of EHBLs at large cosmological redshifts provide an important tool for the study of the history of the Universe through EBL. 
 The hard intrinsic spectrum of EHBL blazars along with the lack of variability also provides a constant VHE flux which is 
 very important for probing the intergalactic magnetic field over a long timescale. The intrinsic VHE photon spectral 
 index $\leq$ 2 indicates a hard underlying particle spectrum in the emission region which challenges the acceleration process 
 in the blazar jet. Due to their hard $\gamma$--ray spectrum, the IC peak in the SED of EHBLs is shifted to TeV energies and this makes 
 their observations with the \emph{Fermi}-LAT at GeV energies very difficult in short time intervals. EHBLs have also been proposed 
 as possible sources of astrophysical neutrinos and ultra-high energy cosmic rays \cite{Padovani2018,Resconi2017}.
 Therefore, the identification of new EHBL candidates is one of the strong motivations for the $\gamma$--ray observations using the upcoming 
 Cherenkov Telescope Array (CTA) observatory and individual systems like H.E.S.S. II and  MACE (Major Atmospheric Cherenkov Experiment) 
 with lower threshold energy below 30 GeV and for the current generation imaging atmospheric Cherenkov telescopes such as MAGIC and VERITAS.   

\section*{Acknowledgements}
We thank the anonymous reviewer for the valuable comments and suggestions that allow us to improve the contents of the manuscript. 
The authors acknowledge the use of the observational results reported in the literature and in the online TeV catalog.

\section*{Appendix: Known EHBLs}

\subsection*{1ES 0229+200}
The blazar 1ES 0229+200 was initially classified as HBL in 1995 from X-ray to radio flux ratio after its discovery in 1992 by the Einstein Slew 
Survey \cite{Giommi1995}. The source is hosted by an elliptical galaxy at redshift $z$ = 0.139 \cite{Woo2005}. The first VHE $\gamma$--ray emission 
from 1ES 0229+200 was discovered by the H.E.S.S. telescopes during 2005--2006 observations above an energy threshold of 580 GeV \cite{Aharonian2007a}. 
The observed VHE spectrum was described by a power law with a spectral index of 2.50$\pm$0.19 in the energy range 500 GeV--15 TeV. The long term 
monitoring of 1ES 0229+200 by the VERITAS telescope during 2009--2012 detected a significant VHE $\gamma$--ray emission from the source with the 
observed spectrum described by a simple power law of spectral index 2.59$\pm$0.12 in the energy range 300 GeV--16 TeV 
\cite{Aliu2014}. The VHE emission detected with the VERITAS telescope indicated a long term variability on a yearly timescale. Near simultaneous 
X-ray observations suggested a hard spectrum with an index $\sim$1.7 and the position of synchrotron peak at energy above 10 keV in the SED. 
The \emph{Fermi}-LAT observations suggested that the high energy spectrum of the source can be described by a hard power 
law with a spectral index of 1.36$\pm$0.25 in the energy range 1--300 GeV \cite{Vovk2012}.

\subsection*{1ES 0347-121}
The blazar 1ES 0347-121 was classified as BL Lac object in 1993 after its discovery in 1992 by the Einstein Slew Survey \cite{Schachter1993}. 
It was observed to harbor a supermassive black hole in the elliptical host galaxy at redshift $z$ = 0.188 \cite{Woo2005}. The H.E.S.S. telescopes 
first discovered a VHE $\gamma$--ray emission from this source in 2006 above an energy threshold of 250 GeV \cite{Aharonian2007b}. The observed 
differential photon spectrum was described by a power law with a spectral index of 3.10$\pm$0.23 and no evidence of short term variability on 
monthly timescales was observed. The \emph{Fermi}-LAT observations indicated high energy $\gamma$--ray emission in the energy range 1-300 GeV to be 
compatible with the power law photon index 1.65$\pm$0.17. The X-ray emission in the energy range 0.3-8 keV was described by a power law model with 
a spectral index of 1.99$\pm$0.06. 

\subsection*{RGB J0710+591}
The blazar RGB J0710+591 was discovered in 1984 as fully resolved elliptical galaxy with a nuclear point source at redshift $z$ = 0.125 \cite{Giommi1991}. 
The first VHE emission from this source was detected by VERITAS during 2008--2009 observations with no evidence of variability in the $\gamma$--ray 
light curve above 300 GeV \cite{Acciari2010}. The observed VHE $\gamma$--ray spectrum was described by a power law with a photon spectral index of 
2.69$\pm$0.26 in the energy range 310 GeV to 4.6 TeV. The HE $\gamma$--ray spectrum in MeV-GeV band was also fitted by a power law with a spectral 
index of 1.46$\pm$0.17 using \emph{Fermi}-LAT observations. The X-ray emission from the source was observed to be consistent with the hard photon 
spectrum having a power law index of 1.86$\pm$0.01 indicating a synchrotron peak energy above 10 keV. The host galaxy of RGB J0710+591 was observed to 
make a significant contribution to the optical emission of the source.

\subsection*{1ES 1101-232}
The blazar 1ES 1101-232 was discovered as a distant VHE $\gamma$--ray source in an elliptical host galaxy at redshift $z$ = 0.186 \cite{Falomo1994}. 
The discovery of VHE $\gamma$--ray emission from this source was reported in 2007 using H.E.S.S. observations performed during 2004--2005 in the 
energy range above 250 GeV \cite{Aharonian2007c}. The observed VHE spectrum was described as a power law function with a spectral index of 
2.94$\pm$0.20 in the energy range of 200 GeV to 4 TeV. No evidence of variability at any timescale in the VHE light curve of the source was detected. 
The X-ray emission of the source in the energy range 3--15 keV was found to be compatible with a broken power law of spectral indices 2.49$\pm$0.02 
and 2.78$\pm$0.16 before and after the break energy ($\sim$ 8 keV) respectively. The peak energy in the first hump of the SED was found to be located 
between 0.5 keV and  3.5 keV. The X-ray and optical emissions from the source were also observed to be constant like their VHE counterpart. 

\subsection*{HESS J1943+213}
The H.E.S.S. telescopes discovered VHE $\gamma$--ray emission from the point source HESS J1943+213 located in the Galactic plane and spatially 
coincident with the hard X-ray unidentified source IGR J19443+2117 during the Galactic plane survey between 2005 and 2008 \cite{Abramowski2011}. 
The time averaged differential VHE spectrum was described by a power law with a soft photon spectral index of 3.1$\pm$0.3 in the energy range 
470 GeV--6 TeV. No significant variability was observed in the VHE $\gamma$-ray emission of the source during this period. The X-ray spectrum of 
the source was observed to be hard with a spectral index $\sim$ 1.04 and without any cut-off up to 195 keV. The IR observations indicated an elliptical 
host galaxy with redshift $z$ $\sim$ 0.14 \cite{Abramowski2011}. The VERITAS observations of this source between May 2014 and November 2015 revealed 
a detection of VHE $\gamma$--rays described by a power law with a spectral index of 2.81$\pm$0.12 in the energy range of 180 GeV to 
2 TeV \cite{Archer2018}. Near simultaneous observation with the \emph{Fermi}-LAT in the energy range 3-300 GeV was used to constrain its 
redshift $z <$ 0.23 \cite{Archer2018}.


\end{document}